# Compositionally Complex Perovskite Oxides as a New Class of Li-Ion Solid Electrolytes

Shu-Ting Ko[1,6], Tom Lee[2,6], Ji Qi[1,6], Dawei Zhang[1,6], Wei-Tao Peng[3,6], Xin Wang[2], Wei-Che Tsai[1], Shikai Sun[2], Zhaokun Wang[2], William J. Bowman[2], Shyue Ping Ong[3,*], Xiaoqing Pan[2,4,5,*], Jian Luo[1,3,7,*]

[1] Materials Science and Engineering Program, University of California San Diego, La Jolla, CA, USA.
[2] Department of Materials Science and Engineering, University of California at Irvine, Irvine, CA, USA.
[3] Department of NanoEngineering, University of California San Diego, La Jolla, CA, USA.
[4] Department of Physics and Astronomy, University of California at Irvine, Irvine, CA, USA.
[5] Irvine Materials Research Institute, University of California at Irvine, Irvine, CA, USA.
[6] These authors contributed equally.
[7] Lead Contact

*Correspondence: jluo@alum.mit.edu (J.L.); xiaoqinp@uci.edu (X. P.); ongsp@eng.ucsd.edu (S.P.O.)

**Summary**

Compositionally complex ceramics (CCCs), including high-entropy ceramics (HECs) as a subclass, offer new opportunities of materials discovery beyond the traditional methodology of searching new stoichiometric compounds. Herein, we establish new strategies of tailoring CCCs via a seamless combination of (1) non-equimolar compositional designs and (2) controlling microstructures and interfaces. Using oxide solid electrolytes for all-solid-state batteries as an exemplar, we validate these new strategies via discovering a new class of compositionally complex perovskite oxides (CCPOs) to show the possibility of improving ionic conductivities beyond the limit of conventional doping. As an example (amongst the 28 CCPOs examined), we demonstrate that the ionic conductivity can be improved by >60% in $(Li_{0.375}Sr_{0.4375})(Ta_{0.375}Nb_{0.375}Zr_{0.125}Hf_{0.125})O_{3-\delta}$, in comparison with the state-of-art $(Li_{0.375}Sr_{0.4375})(Ta_{0.75}Zr_{0.25})O_{3-\delta}$ (LSTZ) baseline, via maintaining comparable electrochemical stability. Furthermore, the ionic conductivity can be improved by another >70% via grain boundary (GB) engineering, achieving >270% of the LSTZ baseline. This work suggests transformative new strategies for designing and tailoring HECs and CCCs, thereby opening a new window for discovering materials for energy storage and many other applications.

# Graphical Abstract

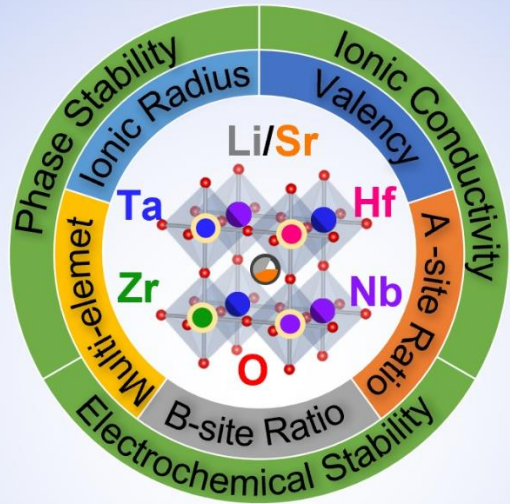
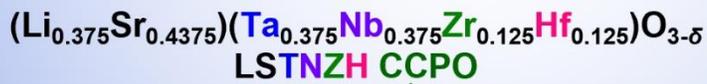
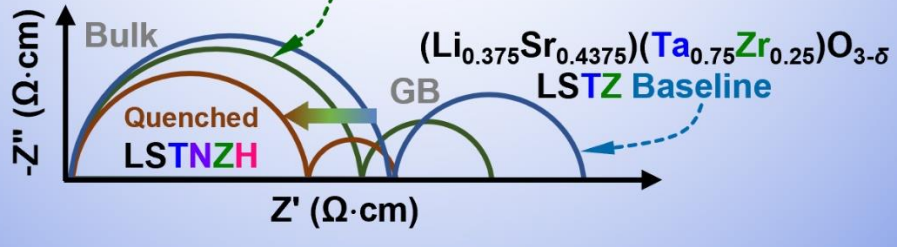

**Introduction**

The classical materials discovery methodology typically relies on searching new stoichiometric compounds, where small amounts of dopants or additives are often introduced to modify and improve the properties. The recent emergence of high entropy ceramics (HECs) of diversifying crystal structures[1–8] unlock new vast compositional spaces with multi-principal (but no dominate) components. Specifically, high-entropy perovskite oxides (HEPOs), which were first reported in 2018,[5] attracted great research interests because of their interesting catalytic,[9] dielectric,[10,11] ferroelectric,[12] magnetic,[13] thermoelectric,[14] magnetocaloric,[15] and electrocaloric[11] properties, as well as promising applications as strongly correlated quantum materials[16] and in solid oxide fuel cells,[17] lithium-ion batteries,[18] and supercapacitors[19]. In 2020, it was further proposed to expand HECs to compositionally complex ceramics (CCCs)[1,20] to consider non-equimolar compositions and short- and long-range ordering that are often essential for further improving and optimizing properties (*e.g.*, the ionic conductivity as we will demonstrate here in this study). Moreover, controlling and tailoring the interfaces and microstructures of HECs and CCCs, along with the non-equimolar compositional designs with aliovalent cations (and vacancies), represent another potentially transformative opportunity that is not yet explored in depth, which motivated this study.

Oxide solid electrolytes are promising candidates for building high energy density all-solid-state batteries (ASSBs), owing to their electrochemical, thermal, and structural stability.[21] Among the different oxide solid electrolytes, perovskite-type $Li_{0.5}La_{0.5}TiO_3$ (LLTO) drew significant attention, owing to its high bulk Li-ion conductivity ($\sigma_b$) in the order of $10^{-3}$ S/cm. Its resistive grain boundaries (GBs), however, constrain its total ionic conductivity for practical applications ($\sigma_{gb}$ ~$10^{-5}$ S/cm so that $\sigma_{total}$ ~$10^{-5}$ S/cm).[22] Furthermore, $Ti^{4+}$ can be reduced to $Ti^{3+}$ at potential below 1.8 V *vs.* $Li/Li^+$,[23,24] which transforms LLTO into an electronic conductor that will ultimately short circuit the battery cell.[25] In attempt to address the shortcomings of LLTO, Chen *et al.* reported perovskite $Li_{0.375}Sr_{0.4375}Ta_{0.75}Zr_{0.25}O_3$ (LSTZ) holding one-order of magnitude higher GB ionic conductivity ($\sigma_{gb}$ ~$10^{-4}$ S/cm) than that of LLTO, a $\sigma_{bulk}$ of $2 \times 10^{-4}$ S/cm, and a wider electrochemical stability window down to 1.0 V *vs.* $Li/Li^+$.[26] LSTZ has shown better performance when compared to LLTO, but its ionic conductivity is still lower than other inorganic solid electrolyte candidates,[27,28] calling for innovative strategies to improve.

In general, $ABO_3$ perovskite oxides have extremely broad applications. It is known that aliovalent doping can tune structures (*e.g.* ordering of A-sites,[29,30] concentration of A-site vacancies,[31] lattice parameter[32,33]) and influence properties including Li-ion conductivity. To date, the majority of the studies have been limited to single- and co-doping. The amount of dopants is typically below 10 mol.% in A- or B-site sublattices to avoid pericipation.[32,34] The versatile structures of perovskites can tolerate a wide range of cation dopants following a criterion using Goldschmidt's tolerance factor ($0.75 < t < 1$),[35] which renders it promising as a model system for exploring high-entropy or complex compositions. Yet, only a single study explored HEPOs as a solid electrolyte, where the reported ionic conductivity was lower than that of baseline material LLTO.[36] In fact, the field of high-entropy solid electrolytes remains largely unexplored, with only few reports published.[36,37,39] Notably, we noticed a most recent study (conducted in parallel with the current study and published just a few days before the submission this article) demonstrated a high-entropy mechanism to improve ionic conductivity, achieving $2.2 \times 10^{-5}$ S/cm in $Li(Ti,Zr,Sn,Hf)_2(PO_4)_3$ as well as two other high-entropy NASICON with ~$10^{-6}$ S/cm ionic conductivities, yet with substantial improvements from their base materials (thereby being promising).[38,40] Herein, we utilized non-equimolar compositional designs to discover a new class of compositionally complex perovskite oxides (CCPOs) to achieve (an order of magnitude higher) $2.56 \times 10^{-4}$ S/cm total ionic conductivity (also representing >270% of the benchmarked state-of-the-art LSTZ baseline, yet with comparable electrochemical stability) with improvements through not only the compositional disorder[38,40] but also microstructure and GB effects.

In this study, we discovered a new class of LSTZ-derived CCPOs as oxide solid electrolytes. We further revealed the phase-microstructure-property relationship to enable us to achieve improved total ionic conductivity, more than doubling of the state-of-the-art LSTZ baseline, in two CCPOs. To investigate the underlying mechanisms, hybrid Monte Carlo/molecular dynamics (MC/MD) and MD simulations using an active learning moment tensor potential (MTP) were conducted to show that no Li depletion at GBs. In addition to potential effects of compositional disorder, we showed that the observed higher total ionic conductivity of CCPOs originate from their larger average grain sizes, which is induced by $Nb^{5+}$ substitution in the B site. Furthermore, we demonstrated that quenching can be used alter GB structures to further enhance the total ionic conductivity, primarily through improving the specific GB ionic conductivity that can be linked to the change in GB structures revealed by advanced microscopy. This work highlights the significance of microstructure and GB controls of CCCs in boosting their ionic conductivities, in addition to non-equimolar composition designs, which points to a new direction to design and tailor inorganic solid electrolytes and potentially many other functional CCCs.

**Results and discussion**

**Discovery of CCPOs and the Composition-Phase-Property Relationship**

All together, we synthesized 28 different compositions of CCPOs in this work through high-energy ball milling and conventional sintering at 1300°C for 12h in air. The key results are summarized in Supplemental Information (Tables S4, S5 and S8).

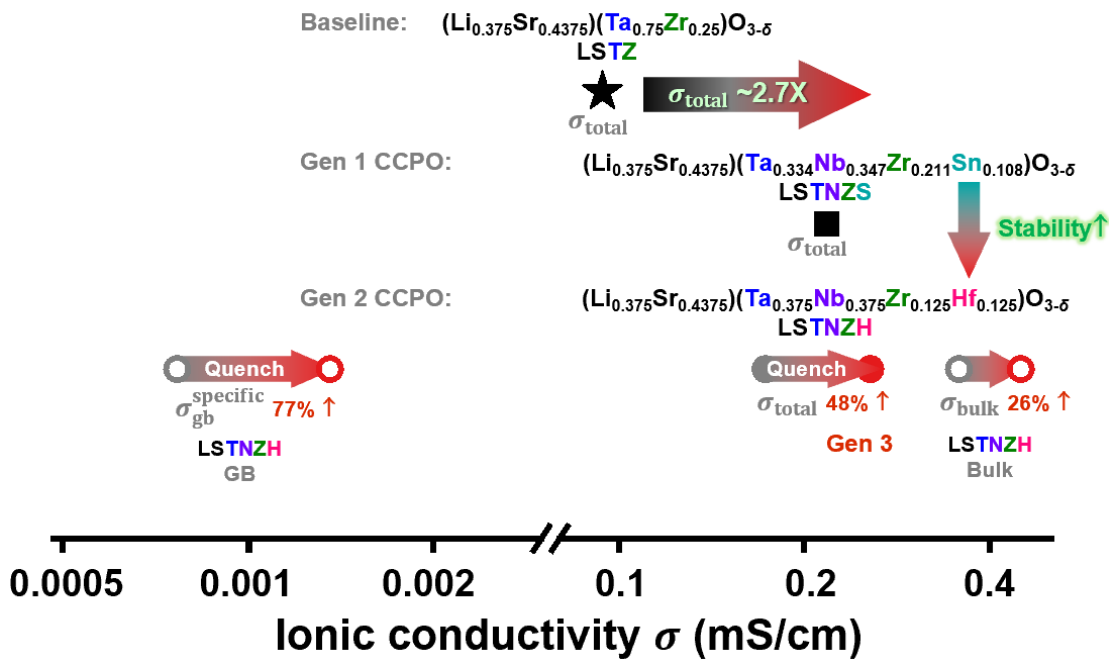

**Figure 1. Overview of several generations of CCPOs discovered and refined in this work, benchmarked with the LSTZ baseline.** The first generation CCPO LSTNZS (square) shows a total ionic conductivity of 0.218 mS/cm, which is ~2.3X of the 0.094 mS/cm of the LSTZ baseline (star), but the electrochemical stability of the Sn-containing LSTNZS is compromised. The second generation Hf-containing (and Sn-free) CCPO LSTNZH (grey circle) shows comparable electrochemical stability with LSTZ and the total ionic conductivity of 0.151 mS/cm, which can be further enhanced through quenching (red circle) to 0.256 mS/cm. The improvement via quenching is primarily due to the increase of specific GB conductivity (by 77%, albeit the bulk conductivity is also improved moderately by 26%). Overall, the total conductivity of the air-quenched LSTNZH is ~2.7X of that of the LSTZ baseline.

Figure 1 presents an overview of several generations of CCPOs discovered and refined in this study, in comparison with the LSTZ baseline. In the first generation, the optimized Sn-containing CCPO ($Li_{0.375}Sr_{0.4375}$)($Ta_{0.375}Nb_{0.375}Zr_{0.125}Sn_{0.125}$)$O_{3-\delta}$ (LSTNZS) shows total ionic conductivity >2.3× higher than that of the LSTZ baseline; however, its electrochemical stability is compromised due to the presence of redox-active Sn (Supplemental Figure S1). In the second generation, Hf-containing (and Sn-free) CCPO ($Li_{0.375}Sr_{0.4375}$)($Ta_{0.375}Nb_{0.375}Zr_{0.125}Hf_{0.125}$)$O_{3-\delta}$ (LSTNZH) shows electrochemical stability comparable with LSTZ with slightly lower ionic conductivity (yet with >60% improvement from the LSTZ baseline), which can be further improved to achieve 0.256 mS/cm total ionic conductivity or 270% of the LSTZ baseline via quenching. This observed enhancement via quenching is primarily from an increase in specific (true) GB ionic conductivity (calculated based on the brick-layer-model described in Supplemental Information Note S1),[41] resulted from changes in the GB compositional profile (segregation) and structure that will be discussed in detail later.

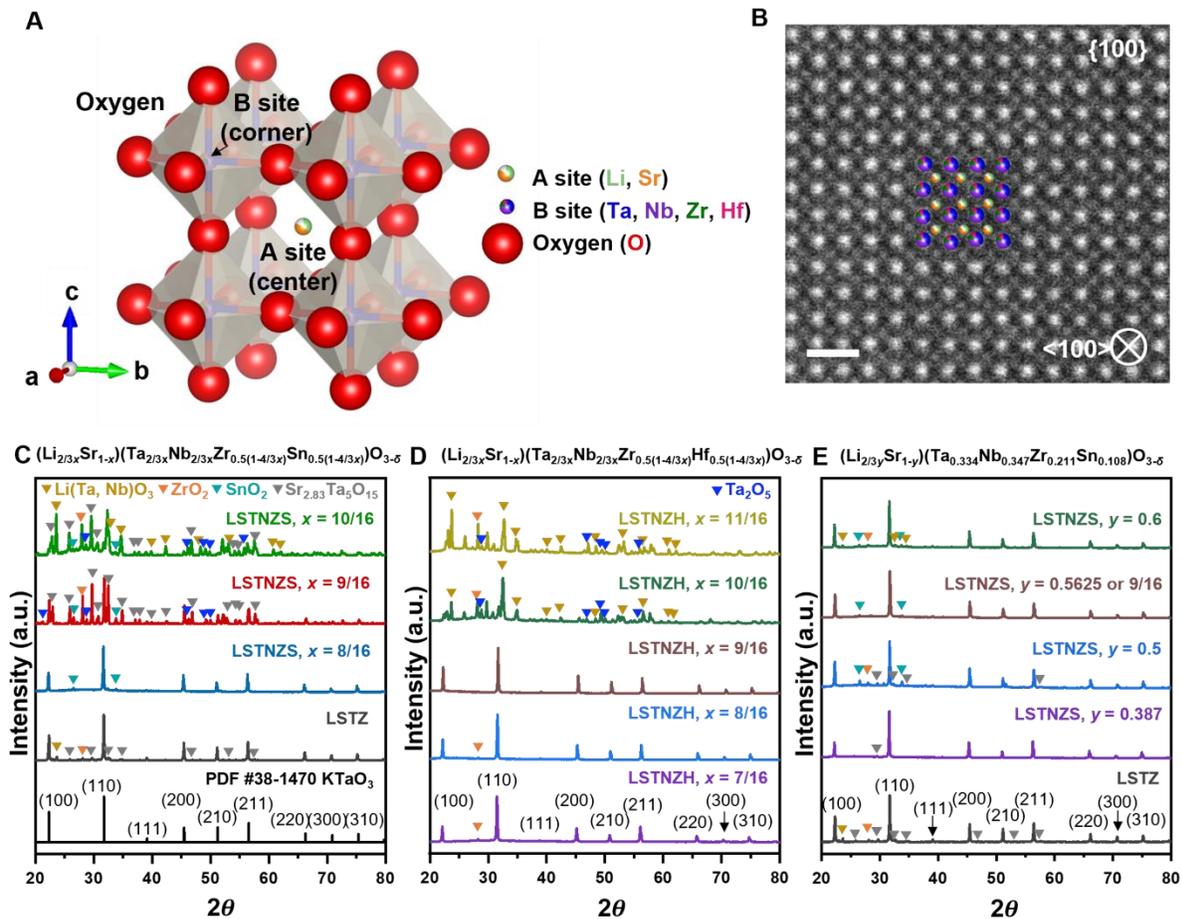

**Figure 2. XRD patterns and crystal structure of new series of CCPO LSTNZS and LSTNZH.** (A) The crystal structure model obtained from XRD refinement of LSTNZH ($x$ = 9/16). The A sites are occupied by Li and Sr cations, while the B-sites are occupied by Ta, Nb, Zr and Sn/Hf cations. (B) The atom arrangement in {100} plane family of LSTNZH ($x$ = 9/16) bulk along <100> under STEM HADAAF imaging (scaler bar = 1 nm). The orange dots indicate A sites, and the blue dots represent B sites. XRD patterns of baseline LSTZ and CCPOs: (C) the LSTNZS $x$ series and (D) the LSTNZH $x$ series, where the stoichiometry controlled by a general formula $Li_{(2/3)x}Sr_{1-x}(5B)_{(4/3)x}(4B)_{1-(4/3)x}O_{3-\delta}$, and (E) the LSTNZS $y$ series with optimized B-site cation ratio (fixed) and changing A-site cation stoichiometry following a formula $Li_{(2/3)y}Sr_{1-y}Ta_{0.3334}Nb_{0.347}Zr_{0.211}Sn_{0.108}$.

For initial design of the CCPO compositions, we adopted a general formula $Li_{(2/3)x}Sr_{1-x}(5B)_{(4/3)x}(4B)_{1-(4/3)x}O_{3-\delta}$ (5B: 5+ cations; 4B: 4+ cations) to control the cation molar ratios, denoted as the "$x$ series". This formula expands on the design formula of simple LSTZ compounds, *i.e.*, $Li_{(2/3)x}Sr_{1-x}Ta_{(4/3)x}Zr_{1-(4/3)x}O_{3-\delta}$,[26] with the assumption of doping $Li^+$ cations into A-site and $Ta^{5+}$ and $Nb^{5+}$ cations into the B-site of the cubic perovskite $Sr(4B)O_3$ framework. Specifically, the A site is occupied by $Li^+$, $Sr^{2+}$, and A-site vacancies. In our initial compositional designs, the 5B are equal molars of $Ta^{5+}$ and $Nb^{5+}$, and the 4B are equal molars of $Zr^{4+}$ and $Sn^{4+}$ for LSTNZS or equal molars of $Zr^{4+}$ and $Hf^{4+}$ for LSTNZH, which are presumably mixed randomly on the B site. Figure 2A displays a schematic of the CCPO unit cell (A-site centered view). Atomic-resolution HAADF-STEM image of the representative CCPO LSTNZH (Figure 2B) and X-ray diffraction (XRD; Figure 2B) confirmed the crystal structure to be cubic perovskite.

First, we investigated the phase formation and stability, which were determined by the amount of primary phase relative to that of secondary phases using XRD. Figures 2C and 2D present the XRD patterns of LSTNZS and LSTNZH $x$ series, respectively, following the $Li_{(2/3)x}Sr_{1-x}(5B)_{(4/3)x}(4B)_{1-(4/3)x}O_{3-\delta}$ formula, where $x$ value is proportional to the total amount of charge carrier ($Li^+$) and A-site vacancies. For LSTNZS ($x = 8/16$), it has a cubic perovskite structure in the space group $Pm\bar{3}m$, which matches the crystal structure of $KTaO_3$ (PDF #38-1470) and that of LSTZ. The phase instability becomes pronounced as the $x$ value increases to 9/16 and higher. At $x \geq 9/16$, $LiNbO_3$-prototyped rhombohedral structure (space group R3c) and $Sr_{2.83}Ta_5O_{15}$-prototyped with tetragonal structure (space group P4/mbm) formed, which consequently triggered the precipitation of secondary $SnO_2$ and $ZrO_2$ phases. The corresponding phase separation is shown in SEM-EDS maps in Supplemental Figure S2. In contrast, the Hf-based LSTNZH system shows a wider single-phase range of up to $x = 9/16$, with only trace amounts of $ZrO_2$ secondary phase found. At $x > 9/16$, the primary phase becomes $LiNbO_3$-prototyped rhombohedral instead of the cubic perovskite phase.

Based on these results, we conclude the single-phase range of CCPOs in the $x$ series is mainly determined by $x$ value, and the instability threshold depends on the difference in cation ionic radii. The LSTNZS $x$ series have a lower phase instability threshold ($x \geq 9/16$) than the LSTNZH $x$ series due to the larger difference in ionic radii between $Sn^{4+}$ and $Zr^{4+}$ (4.17%) than that between $Hf^{4+}$ and $Zr^{4+}$ (1.39%), calculated using Shannon ionic radii and summarized in Supplemental Table S1.[42] This large difference makes it more difficult for LSTNZS to form stable cubic perovskite solid solution at $x \geq 9/16$ and results in precipitation of the secondary $SnO_2$ and $ZrO_2$ phases. Notably, LSTNZS ($x = 9/16$) was benchmarked with Nb and Sn co-doped LSTZ counterparts (LSTZ-NS series) with light doping levels in Supplemental Figure S3. The poor single-phase formability of LSTNZS ($x = 9/16$) illustrates the difficulty in increasing dopant concentration when the large ionic radii difference exists.

To further improve the single-phase formability of the LSTNZS system, we adopted a "natural selection" composition optimization strategy described in the Supplemental Information (Figures S4-S6, Tables S2 and S3) to intentionally synthesize the composition of the primary phase guided from scanning electron microscopy (SEM) energy dispersive X-ray spectroscopy (EDS) quantification results of the single-phase region of the prior generation. Accordingly, a LSTNZS "$y$ series" with the optimal B-site cation ratio, $(Li_{2/3y}Sr_{1-y})(Ta_{0.334}Nb_{0.347}Zr_{0.211}Sn_{0.108})O_{3-\delta}$, was developed. XRD patterns of the LSTNZS $y$ series (Figure 2E) exhibits improved single-phase formability when compared to the LSTNZS $x$ series. For LSTNZS ($y = 0.6$), $LiNbO_3$ is the main secondary phase, where the ratio of A-site cation vacancies to B-site cation vacancies ($[V_A]/[V_B]$) is close to 0.2, resembling that of $A^+B^{5+}O_3$.

Supplemental Figure S7 displays SEM-EDS maps of the LSTNZS $y$ series and LSTNZH $x$ series. To better quantify the secondary phases, Figures 3A and 3B plot secondary phase fractions, which were measured by the areal fractions in the SEM-EDS maps, against compositions of the Sn-containing LSTNZS $y$ series and Hf-containing LSTNZH $x$ series, respectively. The summary of the design workflow of $x$ and $y$ series is illustrated in Supplemental Figure S8. In this study, subsequent characterizations were focused on the Sn-containing LSTNZS $y$ series and Hf-containing LSTNZH $x$ series with better single-phase formability.

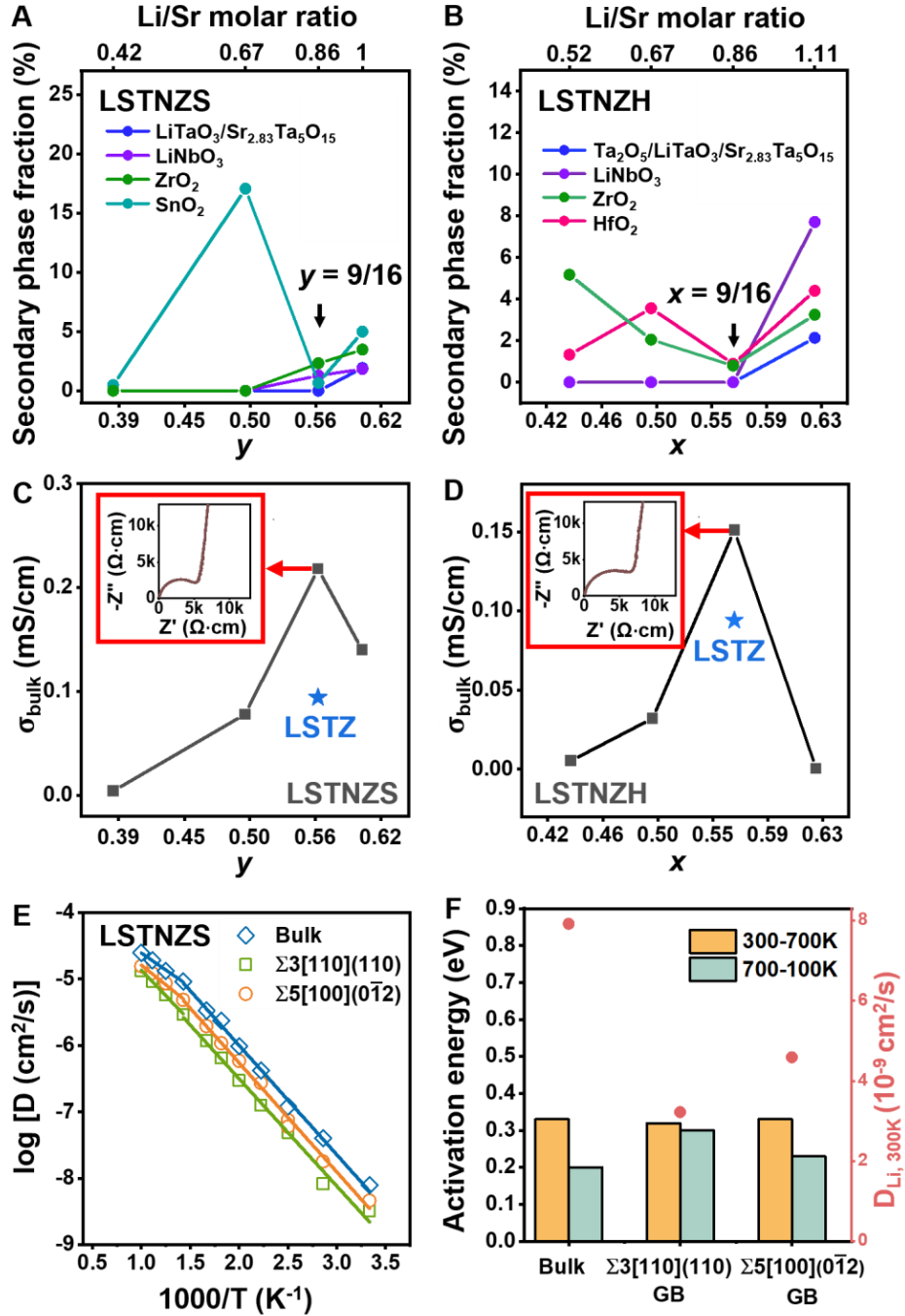

**Figure 3. Plots of secondary phase fraction and bulk ionic conductivity for the LSTNZS $y$ series [(Li$_{2/3y}$Sr$_{1-y}$)(Ta$_{0.334}$Nb$_{0.347}$Zr$_{0.211}$Sn$_{0.108}$)O$_{3-\delta}$] and the LSTNZH $x$ series [(Li$_{2/3x}$Sr$_{1-x}$)(Ta$_{2/3x}$Nb$_{2/3x}$Zr$_{0.5(1-4/3x)}$Hf$_{0.5(1-4/3x)}$)O$_{3-\delta}$] and simulated Li-ion diffusivities in bulk and two grain boundary (GB) models for LSTNZS.** (A) and (B) indicate the areal fraction of the secondary phases, quantified from elemental EDS maps, as function of $y$ or $x$ values (bottom x axis) and the molar ratio of Li/Sr (top x axis). (C, D) Correlation plots of fitted $\sigma_{bulk}$ with $y$ or $x$ values. The $\sigma_{bulk}$ is optimized when $y$ or $x$ = 9/16 or 0.5625 for both the Sn-containing LSTNZS $y$ series and the Hf-containing LSTNZH $x$ series. The total conductivities of two series are benchmarked with the LSTZ baseline (star). Insets are Nyquist plots. (E) The Arrhenius plot of Li diffusivity calculated using the bulk model and two selected LSTNZS ($y$ = 9/16) GBs equilibrated with MC/MD simulations at 1573 K. (F) The corresponding $D_{Li, 300K}$ (right y axis) and activation energies (left y axis).

Second, we evaluated the Li-ion conductivities of the as-synthesized CCPOs. Figure 3C illustrates the influence of Li$^+$ and [$V''_{Sr}$] (in the Kröger–Vink notation) concentrations on the bulk ionic conductivity $\sigma_{bulk}$ (fitted from the measured impedance spectra by the model discussed in Supplemental Note S1) of the Sn-containing LSTNZS $y$ series. When $y$ = 0.387, $\sigma_{bulk}^{LSTNZS}$ is 0.004 mS/cm (the lowest), even though it shows the least fraction of secondary phases (Figure 3A). As $y$ increases to 0.5, the bulk ionic conductivity of LSTNS $\sigma_{bulk}^{LSTNZS}$ is improved by one order to 0.078 mS/cm, even with ~17% SnO$_2$ secondary phase (by volume). Hence, the change of $\sigma_{bulk}^{LSTNZS}$ is dominant by A-site carrier and vacancy amount (rather than phase purities) at $y$ < 9/16 region. Conversely, the phase stability influence becomes dominant when $y \geq$ 9/16. Although the nominal A-site carrier and vacancy amount are the highest at $y$ = 0.6 in Figure 3C, the existence of Sr$_{2.83}$Ta$_5$O$_{15}$, LiTaO$_3$ and LiNbO$_3$ secondary phases indicates the actual Li$^+$ in the main phase is less than the nominal amount. Therefore, the $\sigma_{bulk}^{LSTNZS}$ value reduces to 0.038 mS/cm at $y$ = 0.6. The maximum of $\sigma_{bulk}^{LSTNZS}$ is 0.218 mS/cm at $y$ = 9/16, which is more ionically conductive than that optimal LSTNZS $x$ series composition ($x$ = 8/16; 0.053 mS/cm).

Given that both samples exhibit similar phase stability, the improvement can be attributed to the higher A-site carrier and vacancy concentration while maintaining cubic perovskite structure. To justify the selection of B-site stoichiometry that gives optimal ionic conductivity, an additional LSTNZS Sn/Zr "$w$ series" was fabricated following the formula (Li$_{0.375}$Sr$_{0.4375}$)(Ta$_{0.334}$Nb$_{0.347}$Zr$_{0.319-w}$Sn$_w$)O$_{3-\delta}$, where $w$ variable controls the Sn cation fraction. The results shown in Supplemental Figures S9A and S9C indicate the improvement of both phase stability and $\sigma_{bulk}^{LSTNZS}$ upon increasing the Sn/Zr fraction until more SnO$_2$ forms. Detailed discussion can be found in Supplemental Information. Table S4 summarizes the results for the three Sn-containing LSTNZS series and confirms $y$ = 9/16 in the $y$ series is the composition that provides the highest ionic conductivity.

Likewise, Figure 3D shows the influence of Li$^+$ and [$V''_{Sr}$] concentrations on the bulk conductivity $\sigma_{bulk}$ of the LSTNZH series. The measured $\sigma_{bulk}^{LSTNZH}$ value increases as $x$ increases from 7/16 to 9/16, indicating an increase in the ionic conductivity with concentration of Li$^+$ and Sr vacancies at the single-phase stable regime. Beyond the single-phase stability threshold at $x$ = 9/16, $\sigma_{bulk}^{LSTNZH}$ decreases again due to the presence of secondary phases shown in Figure 3B. Similar to LSTNZS, the optimal A-site ratio takes place at the value where the total A-site carrier and vacancy amount equals 9/16. Table S5 displays the properties for the LSTNZH series discussed above.

Table 1 provides a summary of compositions, sintering conditions, and total ionic conductivity values of representative samples (of optimized compositions) in the present work compared to the results of LSTZ-related compounds reported in the literature.[26,43–45,46]

Complementary to Nyquist plots, we further conducted the distribution of relaxation time (DRT) analysis to deconvolute RC circuit components (polarization processes) involved.[47–50] According to the overlay of DRT peaks with Bode plots in Supplemental Figure S10, CCPOs show polarization processes (P$_1$ and P$_2$) at only high frequency regime (10$^6$ Hz), which are assigned to the responses from bulk. In contrast, additional polarization response is presented in the LSTZ at low frequency regime (10$^4$ Hz), which may come from GB or resistive secondary phase components that requires further investigation. It is worth noting that only one semicircle is shown on the Nyquist plot for these CCPOs (Figures 3C and 3D; Supplemental Figure S11) and no DRT peaks exists at the low frequency regime. Thus, the GB contribution is not the determining factor constraining the total ionic conductivity (a desirable feature), which is different from the case in the well-known Li$_{0.5}$La$_{0.5}$TiO$_3$ ($\sigma_{gb} \approx \sigma_{total} \approx$ 10$^{-5}$ S/cm).[22]

To probe the bulk and GB contributions to the Li diffusivity, we fitted a moment tensor potential for LSTNZS using an active learning strategy and performed MD simulations on three equilibrated structures, *i.e.*, bulk, twist Σ3 [110](110) GB and symmetric tilt Σ5 [100](0$\bar{1}$2) GB to obtain the Arrhenius plots in Figure 3E. It can be observed that regardless of the temperature, the magnitude of $D_{Li}$ follows the order: bulk > simple twist Σ3 [110](110) GB > symmetric tilt Σ5 [100]($0\bar{1}2$) GB. The simulated bulk ionic conductivity at 300 K is 0.285 mS/cm, matching well with the experimentally measured value of 0.218 mS/cm. Also, the activation energy data were summarized in Figure 3F and Supplemental Table S6. A slightly lower activation energy occurs at and above 700 K, comparable with the non-Arrhenius behavior observed in the LSTZ reference.[51] Noticeably, $D_{Li}$ at low-Σ GBs is in the same order of magnitude as $D_{Li}$ in bulk, which matches with the reference LSTZ and differs significantly from that in LLTO. The origin of less resistive GBs of LSTZ has been reported in prior studies.[51] It is worth mentioning that the Li diffusivity at GB and bulk of LSTNZS ($y = 9/16$) are similar to that of LSTZ, which cannot fully explain the unmeasurable GB resistance of LSTNZS ($y = 9/16$) in Nyquist plots. Regardless of GBs, the difference in microstructure between LSTZ and CCPOs likely plays a role in the observed ionic conductivity enhancement.

**The Impact of Microstructures and GB Structures on Ionic Conductivities**

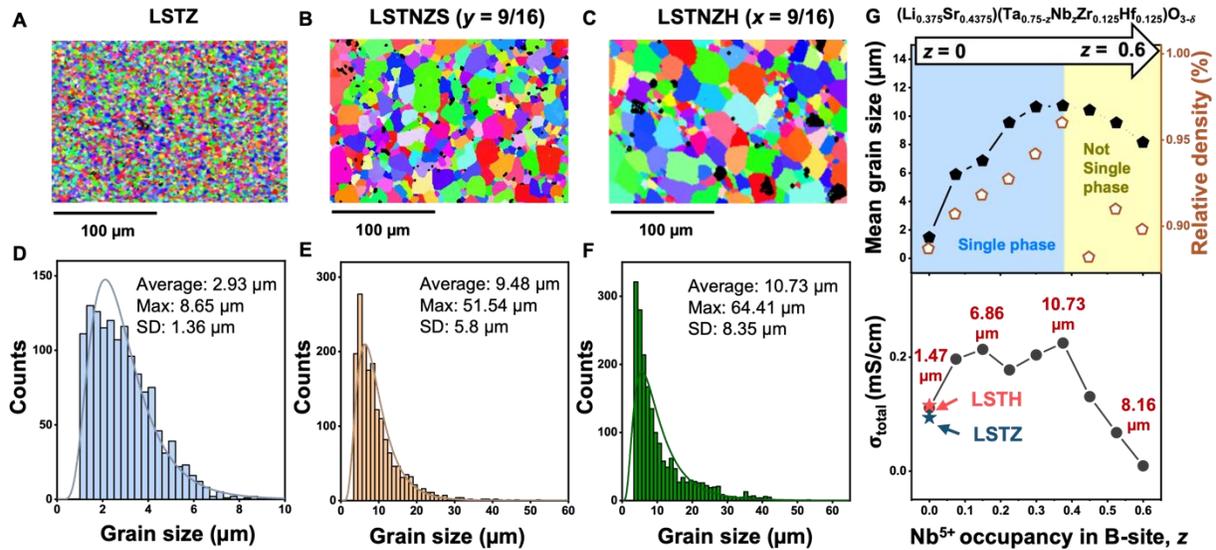

**Figure 4. The inverse pole figures of EBSD, grain size statistics of LSTZ, LSTNZS ($y = 9/16$), and LSTNZH ($x = 9/16$), and the correlation plots of mean grain size and total ionic conductivity with the Nb$^{5+}$ occupancy.** The inverse pole figures along the normal direction of (A) LSTZ, (B) LSTNZS ($y = 9/16$), and (C) LSTNZH ($x = 9/16$) pellets. The corresponding grain size distribution histograms of (D) LSTZ, (E) LSTNZS ($y = 9/16$), and (F) LSTNZH ($x = 9/16$), extracted from (A)-(C), respectively. (G) The correlation of mean grain size (top left y axis), relative density (top right y axis), phase stability (single phase with bule shade; not a single phase with yellow shade), and total ionic conductivity (bottom left y axis) with the Nb$^{5+}$ occupancy [$z$ in (Li$_{0.375}$Sr$_{0.4375}$)(Ta$_{0.75-z}$Nb$_z$Zr$_{0.125}$Hf$_{0.125}$)O$_{3-\delta}$] in B site.

To investigate microstructures of the CCPOs, electron backscattered diffraction (EBSD) in SEM was used to obtain information on grain sizes and orientations. Figures 4A, 4B, and 4C present the orientation maps of representative regions in the LSTZ, LSTNZS ($y = 9/16$), and LSTNZH ($x = 9/16$) pellets along with their corresponding grain size distribution statistics (Figure 4D-F). The color-coded inverse pole figures (Figure 4A-C) indicate both LSTZ and the CCPOs experienced isotropic grain growth and the resulting pellets consisted of largely randomly oriented grains (Supplemental Figure S12A). However, both LSTNZS ($y = 9/16$, ~9.5 μm) and LSTNZH ($x = 9/16$, ~10.7 μm) have average grain size that are 3 times greater than that of the LSTZ reference (~2.9 μm). Additionally, their larger standard deviations (SD) translate to a broader distribution in grain sizes, which correspond to larger kurtosis (sharpness) and skewness (asymmetry) values. Notably, the maxima of grain sizes are ~51.5 μm (LSTNZS) and ~64.4 μm (LSTNZH) in CCPOs, which indicate faster grain growth. The grain size enlargement effectively reduces the overall volume fraction of GBs. As a result, the total ionic conductivities are primarily determined by the bulk conductivity for CCPOs. Thus, only a single semicircle is observed in the Nyquist plots of measured impedance spectra (Figures 3C and 3D).

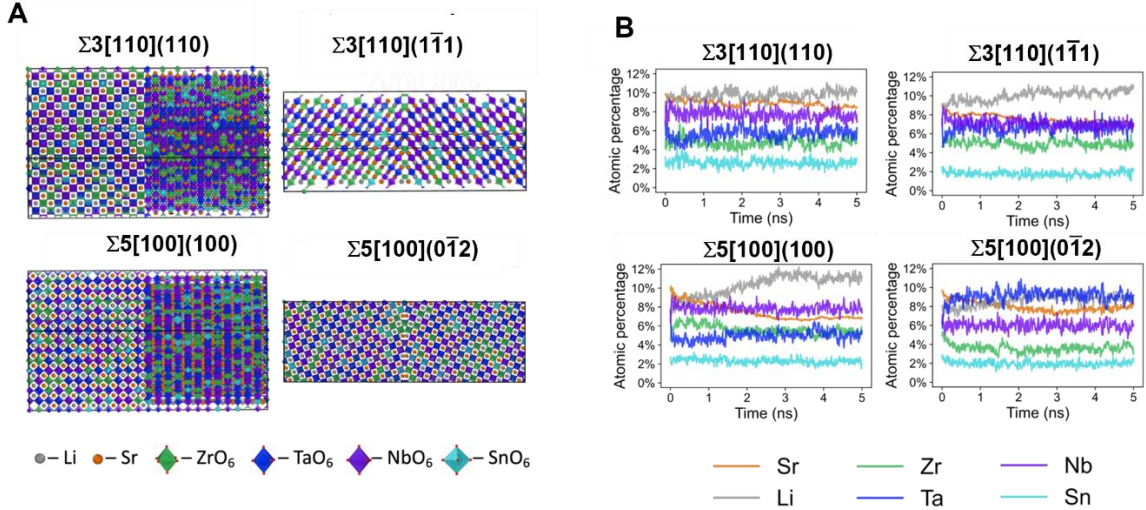

**Figure 5. MC/MD simulated grain boundary (GB) structures of LSTNZS ($y$ = 9/16) based on active learning MTP.** (A) Four GB structures of LSTNZS ($y$ = 9/16) before MC/MD simulations were constructed with large-scale GB models with over 10,000 atoms. Elements on A- and B-sites are randomly assigned. The exact geometric information of those GB models was provided in Supplemental Table S7. (B) The evolution of the atomic percentage of cations at GB regions of the four GB models during the MC/MD simulations at 1573 K.

To further probe the origin of larger grain sizes in CCPOs, an additional "$z$ series" of LSTNZH $(Li_{0.375}Sr_{0.4375})(Ta_{0.75-z}Nb_zZr_{0.125}Hf_{0.125})O_{3-\delta}$, where only the Ta/Nb ratio was varied and all other cations are kept to the same stoichiometry as that in LSTNZH ($x$ = 9/16), were synthesized and characterized. Figure 4G plots the measured mean grain size, relative density, and total ionic conductivity of $(Li_{0.375}Sr_{0.4375})(Ta_{0.75-z}Nb_zZr_{0.125}Hf_{0.125})O_{3-\delta}$ against $z$ (*i.e.*, the Nb fraction on the B site). In general, all three parameters (grain size, relative density, and total ionic conductivity) increase with the Nb fraction $z$ up to $z$ = 9/16. At $z$ > 9/16, a single phase is no longer maintained, and all three parameters generally decrease with the further increasing Nb fraction $z$ . The characterization of phase stability, grain size distribution, and ionic conductivity are summarized in Supplemental Figures S13-S16. These results suggest that $Nb^{5+}$ substitution in B-site promotes grain growth and broadens the grain size distribution, coincident with the ionic conductivity enhancement until the single-phase stability threshold. The large grains are homogeneously distributed throughout the samples, as shown in Supplemental Figure S14. $Nb_2O_5$ is known as a sintering additive and has been reported to promote grain growth.[52,53] In fact, single crystals are favorable for improving total ionic conductivity and enhancing material rigidity by completely eliminating GBs.[54,55] However, the cost of growing single crystals makes this approach impractical for large-scale manufacturing. Here, our series of $(Li_{0.375}Sr_{0.4375})(Ta_{0.75-z}Nb_zZr_{0.125}Hf_{0.125})O_{3-\delta}$ demonstrates a simple, cost-effective way to decrease volume fraction of GBs and increase ionic conductivity via the addition of $Nb_2O_5$ to promote grain growth.

Apart from the microstructure observations, Supplemental Figures S17A and S17B display atomic-resolution high-angle annular dark-field scanning transmission electron microscopy (HAADF-STEM) images of LSTNZS ($y$ = 9/16) general GB and (010) faceted GBs (with respect to the left-side grain), respectively. The dark bands observed at both GBs can indicate compositional or mean density variation across the GB. Hence, STEM-EELS measurements were performed to unravel the change of compositions at GBs. However, owing to the lower volume densities of Li$^+$ ions compared to other types of electrolytes and the poor scattering power of Li ($Z$ = 3),[46,47] the Li-K edges cannot be detected in the STEM EEL spectrum shown in Figure S19A. The more details are discussed in Supplemental Note S3.

To further investigate the local composition of LSTNZS ($y$ = 9/16) GBs beyond experimental limitations, we have fitted an active learning MTP potential to simulate and compare bulk and GB structures of LSTNZS ($y$ = 9/16). Our MTP is verified to have excellent accuracy in reproducing density functional theory (DFT) energies, forces, stresses, and GB energies (see Supplemental Note S5). With this MTP, hybrid MC/MD simulations were conducted at the experimental calcination temperature of 1573 K. The evolution of GB composition is shown in Figure 5B and Table 2. In line with our previous computational results in the LSTZ reference,[51] A-site Sr vacancies and preserved or increased Li concentration were generally observed at the GB regions of LSTNZS ($y$ = 9/16). The absence of Li depletion at GB regions, which is observed in the resistive GB of the LLTO, promotes Li diffusion at GB regions. In terms of B-site composition, the equilibrated atomic percentage at GB is generally comparable to that of bulk, while slight variations are found to be GB orientation dependent. For instance, the atomic percentage of Ta atoms increased from 6.4% to 9.2%, which is the highest among all other elements at the symmetric tilt Σ5 [100]($0\bar{1}2$) GB. On the other hand, Nb atoms become enriched at the simple twist Σ5 [100](100) GB, whereas the atomic percentage of Ta decreases from 7.8% to 5.4%. This result indicates the variations in B-site compositions at GB regions at 1573 K. Therefore, the elemental segregation observed in STEM-EELS map in Supplemental Figure S18C can be a phenomenon depending on the GB character (crystallographic anisotropy).

**Conductivity Improvements via Quenching for LSTNZH**

Having improved the ionic conductivities of CCPOs to higher than those of the baseline LSTZ, we further demonstrated the performances (conductivity) of CCPOs can be further tuned and improved via employing different cooling rates, which can substantially improve the total conductivity of the LSTNZH ($x$ = 9/16), primary through changing the GB segregation (compositional profiles) and structure and increasing the specific GB conductivity. This opens a new window to tailor and improve CCPOs and potential other solid electrolytes.

On the one hand, Figure S20A displays the Nyquist plots of furnace-cooled (brown curve) and air-quenched (orange curve) Sn-containing LSTNZS ($y$ = 9/16). For the air-quenched sample, a second arc attributed to polarization of GBs, secondary phase, and defects has appeared; the calculated bulk, GB, and total ionic conductivity ($\sigma_{bulk}$, $\sigma_{gb}$, and $\sigma_{total}$, respectively, following Supplemental Note S1) are $\sigma_{bulk}$ = 2.37 × 10$^{-4}$ S/cm, $\sigma_{gb}$ = 3.90 × 10$^{-4}$ S/cm, and $\sigma_{total}$ = 1.47 × 10$^{-4}$ S/cm. When comparing the XRD patterns of the furnace-cooled and air-quenched sample (Supplemental Figure S20B), the latter shows more pronounced ZrO$_2$ and SnO$_2$ peaks as well as appearance of the SrO peaks. SEM-EDS data (Supplemental Figure S20C) further confirms the increase in the amounts of the secondary phases in the air-quenched LSTNZS ($y$ = 9/16). In summary, quenching generated additional secondary phases and caused the total conductivity $\sigma_{total}$ to decrease (by ~32.6%) in Sn-containing LSTNZS, which is undesirable.

On the other hand, similar experiments and characterization were performed on the Hf-containing LSTNZH ($x$ = 9/16) samples to show a beneficial effect of quenching. Figure S20D shows the Nyquist plots of furnace-cooled (gray curve) and air-quenched (red curve) pellets. A smaller arc measured from the latter indicates quenching LSTNZH ($x$ = 9/16) resulted in a desirable increase in the total conductivity $\sigma_{total}$. When comparing data of the air-quenched sample to that of furnace-cooled, XRD pattern (Figure S20E) and SEM-EDS analysis (Figure S20F) both confirm that quenching LSTNZH ($x$ = 9/16) does not generate additional secondary phases. The results thus suggest that (Li$_{2/3x}$Sr$_{1-x}$)(Ta$_{2/3x}$Nb$_{2/3x}$Zr$_{0.5(1-4/3x)}$Hf$_{0.5(1-4/3x)}$)O$_{3-\delta}$ has a greater stability threshold than both (Li$_{2/3x}$Sr$_{1-x}$)(Ta$_{2/3x}$Nb$_{2/3x}$Zr$_{0.5(1-4/3x)}$Sn$_{0.5(1-4/3x)}$)O$_{3-\delta}$ and $(Li_{2/3y} Sr_{1-y})(Ta_{0.334}Nb_{0.347}Zr_{0.211}Sn_{0.108})O_{3-\delta}$. This is likely due to the smaller difference in ionic radii between $Hf^{4+}$ and $Zr^{4+}$ (1.39%) than that between $Sn^{4+}$ and $Zr^{4+}$ (4.17%).

Comparing Sn-containing LSTNZS and Hf-containing LSTNZH, we can conclude that the choice of elements, stoichiometric ratio, differences in ionic radii, and overall valency concurrently influence the phase stability and consequently macroscopic property. It is possible to achieve desirable conductivity improvement via quenching (through the improvement in the specific GB conductivity as we will show next) if the primary phase is sufficiently stable (*e.g.*, in Hf-containing LSTNZH).

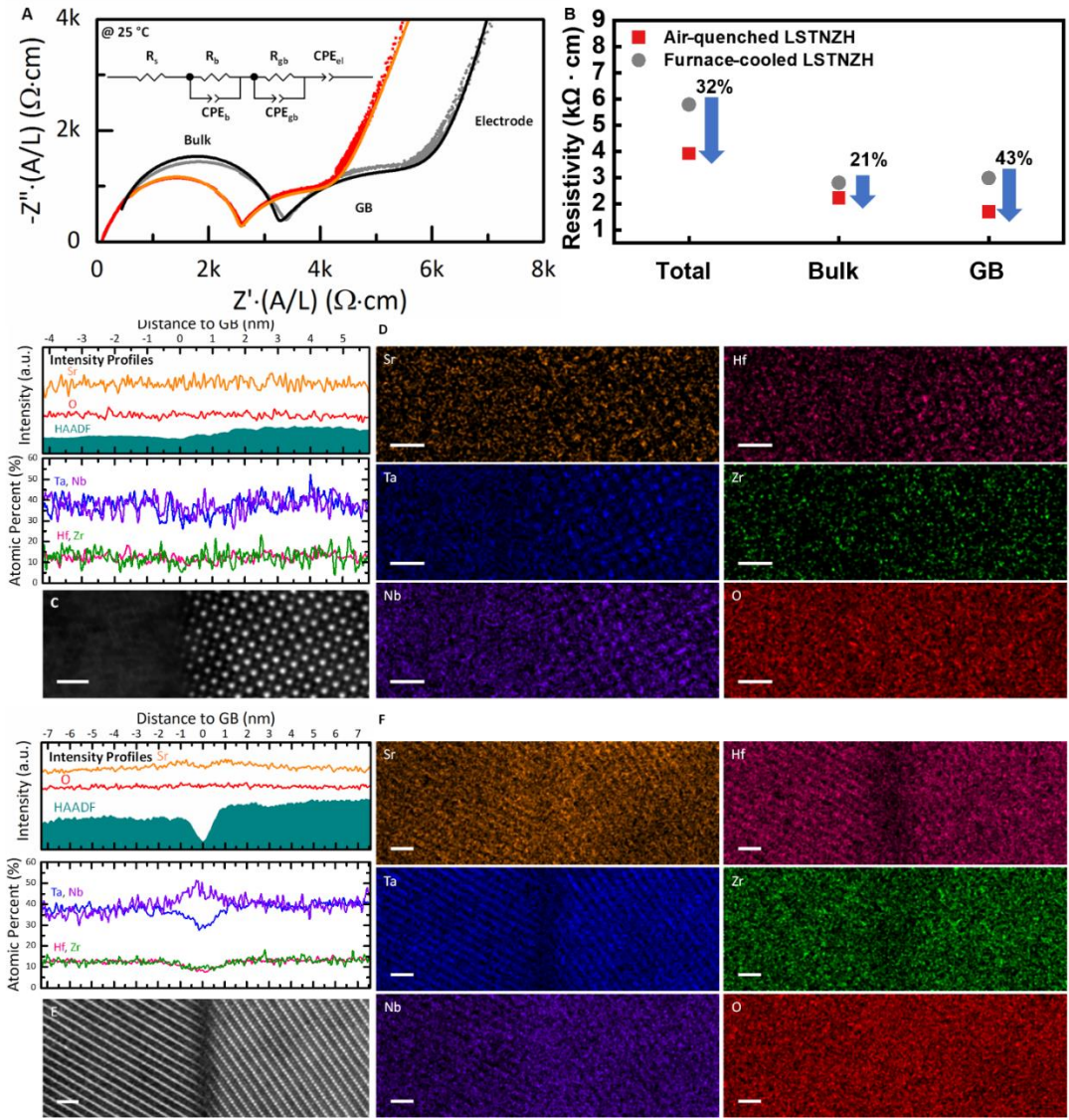

**Figure 6. The effects of the cooling rate on structure-property relationship of LSTNZH (*x* = 9/16) samples, including GB structure (segregation) and conductivity.** (A) AC impedance spectra of furnace-cooled (gray curve) and air-quenched (red curve) LSTNZH (*x* = 9/16) samples measured at 25 °C. Equivalent circuit model used to fit the data is shown in (A). (B) Bulk, GB, and total resistivity of furnace-cooled and air-quenched LSTNZH (*x* = 9/16) samples. (C) HAADF-STEM image and (D) elemental maps of Sr, Ta, Nb, Hf, Zr, and O collected at the furnace-cooled LSTNZH (*x* = 9/16) general GB. (E) HAADF-STEM image and (F) elemental maps of Sr, Ta, Nb, Hf, Zr, and O collected at the air-quenched LSTNZH (*x* = 9/16) general GB. Scale bars are 1 nm. Additional examples are given in Supplemental Information to show the statistical significance of the observations.

To better understand the influence of cooling rate to the conductivity improvement of the LSTNZH ($x$ = 9/16), larger pellets were synthesized to give greater number of GBs, so that the polarization of the GB component is more pronounced in the AC impedance measurement. Figure 6A shows the Nyquist plots of furnace-cooled (gray curve) and air-quenched (red curve) LSTNZH ($x$ = 9/16). For the air-quenched sample, the calculated bulk, (apparent) GB, and total resistivity ($\rho_{bulk}$, $\rho_{gb}$, and $\rho_{total}$, respectively) are $\rho_{bulk} \approx 2.22 \times 10^3$ Ω·cm, $\rho_{gb} \approx 1.69 \times 10^3$ Ω·cm, and $\rho_{total} \approx 3.91 \times 10^3$ Ω·cm, which are ~21%, ~43%, and ~32% lower than those of the furnace-cooled sample (Figure 6B). The corresponding bulk, (apparent) GB, and total ionic conductivities of the air-quenched LSTNZH ($x$ = 9/16) are $\sigma_{bulk} \approx 0.45$ mS/cm, $\sigma_{gb} \approx 0.59$ mS/cm, and $\sigma_{total} \approx 0.256$ mS/cm, which represents ~26%, ~77%, and ~48% increase from the furnace-cooled sample (as shown in Figure 1, with data listed in Supplemental Table S8). Moreover, the specific (true) GB conductivities were calculated to be $\sigma_{gb}^{specific} \approx 1.36 \times 10^{-6}$ S/cm for the air-quenched LSTNZH vs. $\sigma_{gb}^{specific} \approx 0.77 \times 10^{-6}$ S/cm for the furnace-cooled LSTNZH ($x$ = 9/16) based on the model approach described in Supplemental Note S1.

**An In-Depth Study of the Cooling Rate Effects on GB Structure and Conductivity**

To understand the cause of this decreased resistivity, high spatial resolution STEM-EDS measurements were performed at general GBs of furnace-cooled vs. air-quenched LSTNZH samples.

Atomic-resolution HAADF-STEM image of a general GB in the furnace-cooled LSTNZH ($x$ = 9/16) is shown in Figure 6C, from which the EDS spectrum was collected. Figure 6D displays the atomic-resolution elemental maps of Sr, Ta, Nb, Hf, Zr, and O. The maps were generated using intensities from Sr-K, Ta-L, Nb-K, Hf-L, Zr-K, and O-K edges. To analyze the correlation between HAADF image and the elemental maps, plots of vertically integrated intensity profiles across the GB are shown above the HAADF image (Figure 6C). The intensity profile of the HAADF image increases from left to right because the right grain is in the low-index [100] zone axis, and HAADF intensity of the GB is close to that of the left grain. Consistent with the HAADF image, intensity profiles of Sr, Zr, Hf, and O are mostly uniform while those of Ta and Nb increase slightly from left to right. Similar results are found at a small-angle GB of furnace-cooled LSTNZH ($x$ = 9/16) (Supplemental Figure S21).

In contrast, striking differences were observed in STEM-EDS of the air-quenched sample. Figure 6E displays the atomic-resolution HAADF-STEM image of an air-quenched LSTNZH ($x$ = 9/16) general GB. Similar to the LSTNZS ($y$ = 9/16) GBs, an obvious dark band is observed, indicating a compositional variation across the air-quenched LSTNZH ($x$ = 9/16) GB. Figure 6F displays the atomic-resolution elemental maps of Sr, Ta, Nb, Hf, Zr, and O generated using the same edges as those used in Figure 6D. To analyze the correlation between HAADF image and the elemental maps, plots of vertically integrated intensity profiles across the GB are again shown above the HAADF image (Figure 6E).

For the Sr profile, a decrease at the GB and an increase at the left and right sides adjacent to the GB was observed. This indicates that the GB has undergone elemental segregation, with Sr segregating out of the GB core and into sides of the abutting grains. Such off-center segregation profile, which has been observed in limited other materials such as Co and Ti co-doped WC,[60] represent an interesting feature showing complex interactions of multiple elements (cations) at the GB.

For the B-site elemental signals, Ta, Zr, and Hf decrease while Nb increases at the GB. This indicates Nb is substituting Ta, Zr, and Hf at the GB. Both Sr elemental segregation and Nb substitution suggest greater chemical disorder at the GBs of air-quenched LSTNZH ($x$ = 9/16), which can be a factor contributing to their decreased resistivity. To further quantify the B-site stoichiometry at the LSTNZH GBs, Supplemental Figures S22A, S22B, and S22C plot the total B-site atomic percent against distance to GB for GBs examined in Figure 6C, Supplemental Figure S21B, and Figure 6E, respectively. Unlike plots for furnace-cooled LSTNZH (Supplemental Figures S22A and S21B), the plot for air-quenched LSTNZH (Supplemental Figure S22C) shows a pronounced decrease in total B-site atomic percent at the GB. This suggests that quenching resulted in increased B-site vacancies at the GBs, and that Nb does not stoichiometrically fully compensate the decrease in Ta, Zr, and Hf. Since the O signals do not show a decrease at GBs, the air-quenched LSTNZH GBs likely have higher Li concentration than the nominal value (or net negative charges). Although the Li content cannot be experimentally determined for LSTNZH as well, owing to distinct Li-K edge not observed in low loss EELS (Supplemental Figure S23), both excess $Li^+$ charge carrier at GBs or net negative charges can decrease the GB resistivity.

Moreover, the decrease in HAADF intensity at GB is consistent with the trends observed in elemental signals. Since HAADF-STEM detects inelastically scattered electrons transmitted through the STEM specimen, having less Sr results in less scattering and therefore produces darker A-site atomic columns at the GB core than at the sides of the abutting grains. Similarly, having B-site vacancies at GBs results in darker B-site atomic columns. Furthermore, atomic columns with higher average atomic number appear brighter in HAADF-STEM images, so lighter Nb ($Z = 41$) substituting heavier Hf ($Z = 72$) and Ta ($Z = 73$) results in darker B-site atomic columns at GB. It is also possible the general GBs in quenched specimens are more disordered (quenching the high-temperature interfacial disordering that has been widely observed[62,64,66] and can be enhanced by compositional complexity[68]). Thus, the like effects of quench are as follows. It causes compositions of both GBs and the nearby bulk regions to deviate from the nominal stoichiometry $(Li_{0.375}Sr_{0.4375})(Ta_{0.375}Nb_{0.375}Zr_{0.125}Hf_{0.125})O_{3-\delta}$. This change in composition likely results in increased B-site vacancies and potentially greater chemical (and structural) disorder at the GBs, all of which can facilitate $Li^+$ ion migration to decrease GB resistivity.

It should be noted that multiple GBs were characterized for both air-quenched and furnace-cooled specimens to ensure the observed differences are general and statistically significant. Additional examples are documented in Supplemental Information. Further STEM-EDS measurements were performed at one GB formed by grains that has average grain size (Supplemental Figure S24) and at two GBs by grains that have above average grain sizes (Supplemental Figure S25). The results are similar to those found in the general GB shown in Figure 6E. From these observations, it is reasonable to infer that Sr segregation and Nb substitution are present in all general GBs of air-quenched LSTNZH, regardless of the sizes of grains forming the boundaries. Following the same analysis performed earlier, Supplemental Figures S22D, S22E, and S22F plot the total B-site atomic percent against distance to GB for the GBs examined in Supplemental Figure S24C, Figure S25C, and Figure S25E, respectively. Both Supplemental Figure S22D and S22F display a pronounced decrease in total B-site atomic percent at the GBs, further providing confidence to the conclusion that air quenching promotes B-site vacancies at GBs. As for Supplemental Figure S22E, the decrease is not observed due to a thickness gradient in this selected region, thinner on left and thicker on right of GB, which might be excluded as an artifact of single incident.

It is worth noting that despite the compositional change, the cubic perovskite crystal structure is still maintained at the GBs of the air-quenched sample, albeit the general GBs can be more disordered as discussed above, as (premelting like) high-temperature GB disordering induced by the temperature and enhanced by segregation may be partially quenched.[62,64,66] HAADF-STEM images (Figure 6E and Supplemental Figures S24C, S25C, and S25E) clearly show the same crystal lattices from the bulk extend all the way to the GBs, with the observed decrease in intensity at GBs. Thus, GBs of air-quenched LSTNZH likely consist of defective perovskite structures with increased number of B-site vacancies and possible (relatively low levels of) GB structural disordering.

Furthermore, density functional theory (DFT) calculations were conducted to investigate the A-site elemental contents at LSTNZH ($x = 9/16$) GBs. Twenty randomly ordered LSTNZH systems that satisfied nominal stoichiometry at both bulk and GB regions were generated for $\Sigma5$ [100]($0\bar{1}2$) and $\Sigma3$ [110]($1\bar{1}1$) GBs (see Figure 7A and Figure 7B for the lowest-energy relaxed structures). After relaxation, the changes of A-site elemental fractions have been observed in the lowest energy structure of $\Sigma5$ [100]($0\bar{1}2$) GB (the change of fraction from 7/16 to 3/16 for Sr, and from 6/16 to 5/16 for Li), demonstrating a higher tendency for Sr ions to leave the GB regions. For each type of GBs, three lowest-lying relaxed structures which represent various local ordering of A-site and B-site elements were selected for further stabilization energy calculations. The stabilization energy is defined as the total energy of the system with a target A-site ion placed at nearby A-site vacancy away from GB center minus the total energy of the original system with the ion located at the GB region (see Figure 7A for an example), which estimates the driving force for the depletion of the ion. As shown in Figure 7C, Sr ions in general possess lower stabilization energies, suggesting a larger driving force to depart from GB regions than Li ions for the LSTNZH system. The result is in line with the MTP MC/MD simulations for the comparable system LSTNZS, where Sr depletion and maintaining Li concentration at GBs have been found. The trend can be attributed to the distinct coordination environments and the shorter A-site to A-site distance at the GB regions, which have a larger effect on the Sr ion due to its larger ionic radius and valency charge. It appears to be consistent with the off-center GB segregation of Sr, with relative Sr depletion at the GB core, revealed by STEM EDS (Figure 6E), although it does not explain the off-center segregation directly, which can be a complex interplay (including possible GB disordering effects) at more general GBs (than the simplified special Σ GBs that can be modeled by DFT). Importantly, the calculations suggest less tendence of Li depletion at the GBs for LSTNZH, so that it is less a concern to reduce the specific GB conductivity of this class of materials (consistent with the MTP MC/MD results of LSTNZS shown above).

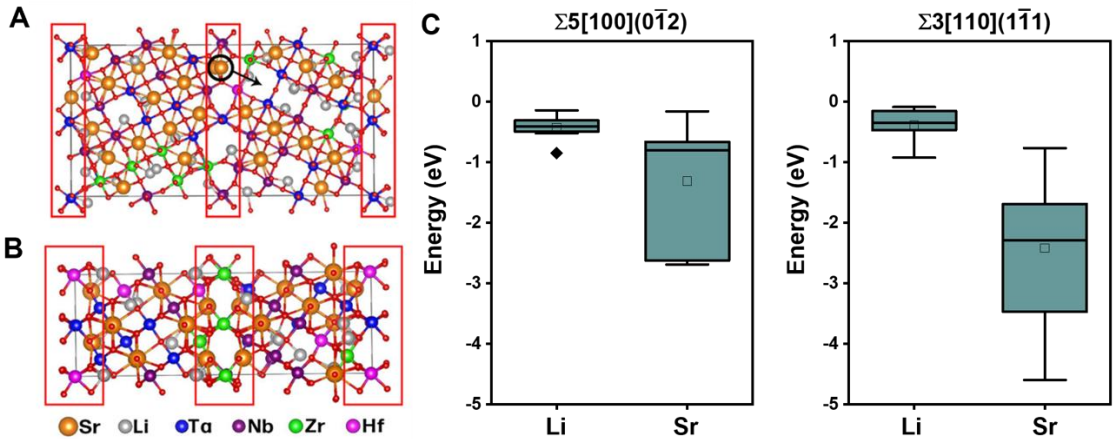

**Figure 7. GB models and the corresponding DFT calculations for the stabilization energy investigation.** (A) and (B) present the lowest-energy relaxed Σ5 [100]($0\bar{1}2$) and Σ3 [110]($1\bar{1}1$) structures of LSTNZH, respectively. The red rectangles indicate the GB regions in the model systems. In (A), the black circle and arrow indicate a Sr locating at GB to be placed at a nearby vacancy away from GB center for calculating the stabilization energy (the driving force of elemental depletion). (C) The result of stabilization energy calculations in box plots. The squares indicate the mean values, and the black diamond represents an outlier.

**Electrochemical Stability**

To characterize the electrochemical stability window of our CCPOs, cyclic voltammetry (CV) scanned at 0.1 mV/s was performed using half-cell configuration. Figure 8A presents an overlay of the cyclic voltammograms to compare the reduction limits of LSTNZS ($y = 9/16$), LSTNZH ($x = 9/16$), and LSTZ. The onset reduction potential, the voltage at which a large amount of reduction reaction begins to take place, is defined as the intersection point of the dashed line extrapolated from the low current response region and that extrapolated from the large negative current response region. The onset reduction potential is 1.4 V for both CCPOs and LSTZ (inset in Figure 8A). On the one hand, LSTNZS displays an additional oxidation peak at 0.5 V, corresponding to Sn/Li dealloying reaction.[58] On the other hand, LSTNZH exhibits reduction and oxidation onset potentials comparable to those of LSTZ despite its compositional complexity. According to the grand potential diagram in Supplemental Figure S1, the calculated electrochemical stability window of LSTNZH is 1.63-3.58 V and that of LSTZ is 1.13-3.56 V. The small discrepancies between simulated and experimental results may be due to sluggish kinetics of decomposition reactions and passivation by decomposition interphase, which was not accounted for in the theoretical calculations.[59]

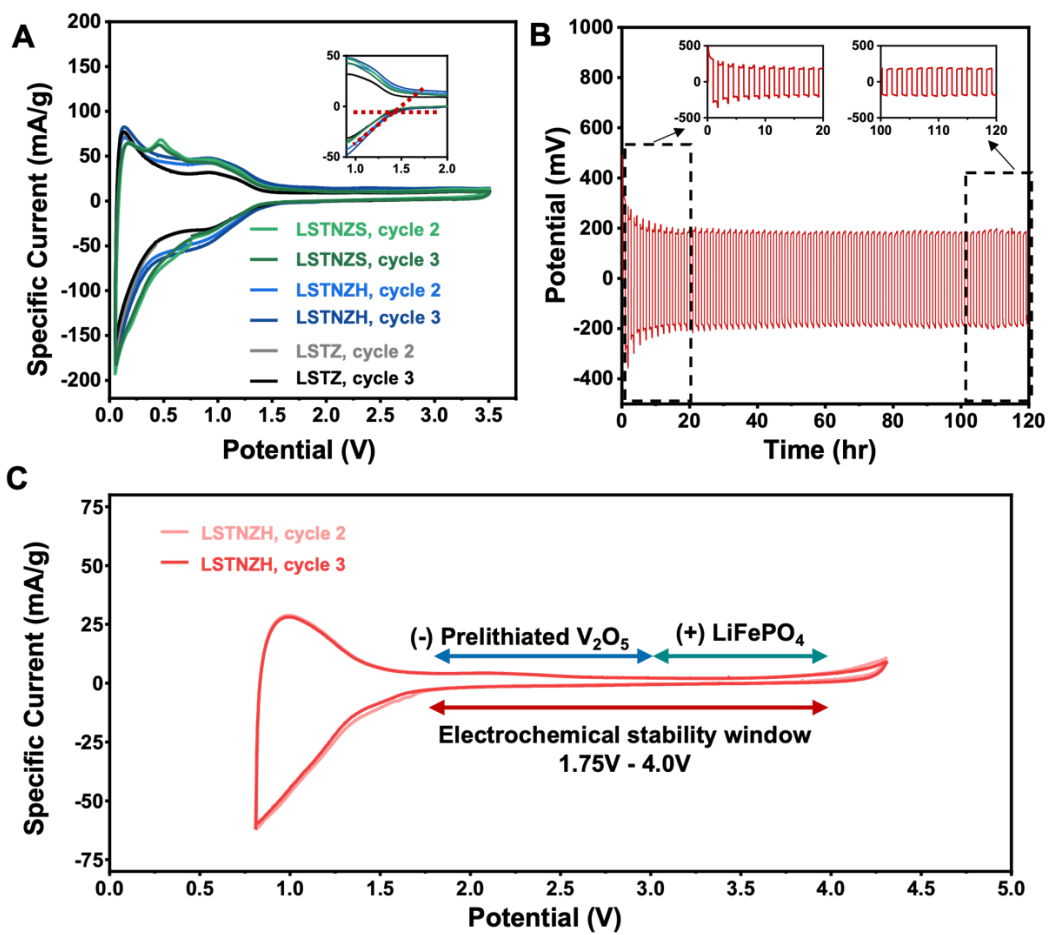

**Figure 8. The electrochemical stability testing by cyclic voltammogram using half cells for LSTNZS, LSTNZH, or LSTZ.** (A) The comparison of CV curves among three sample groups, as indicated: LSTNZS (green), LSTNZH (blue), and LSTZ (grey). The half cells were assembled using Li metal (anode), LP40 with 10 vol% FEC (liquid electrolyte), and testing sample coated on Cu foil (cathode). The CV measurements were scanned within the potential window of 0 V - 3.5 V, with a scan rate of 0.1 mV/s. The enlarged region shows the onset reduction potential at 1.4 V. (B) shows the Li-Li symmetrical cell testing using LSTNZS as the solid electrolyte separator with trace LP40 liquid electrolyte and 3501 Celgard separators on both sides between LSTNZS and Li metal. The cell was tested at 300 µA cm$^{-2}$ for 120 h, with an overpotential of 200 mV. (C) Cyclic voltammogram of half-cell assembled using Li metal (anode), LP40 with 10 vol% FEC (liquid electrolyte), and LSTNZH coated on Al foil (cathode). The CV curves were scanned from 0.8 V to 4.3 V at a scan rate of 0.1 mV/s.

To validate the cyclability of CCPOs, the liquid-solid hybrid Li | LSTNZS | Li symmetrical cell was assembled and tested at 300 µA cm$^{-2}$. The result in Figure 7B shows that the cell was cycled for 120 hours with a stable overpotential of 200 mV for Li stripping/plating reactions before polarization increased. Since LSTNZH is more electrochemically stable than LSTNZS, CV of LSTNZH coated on Al foil was measured to further test its stability at high voltage. Cyclic voltammogram shows stable cycling from 1.75

V to 4.0 V when the electrode was evaluated from 0.8 V to 4.3 V (Figure 8C). Given this electrochemical stability window, one possible ASSB would be LiFePO$_4$ ∥ LSTNZH ∥ prelithiated V$_2$O$_5$. The voltage profiles in Figure S26 show the galvanostatic charge and discharge curves of prelithiated V$_2$O$_5$ and LiFePO$_4$ half-cells, demonstrating their working windows fall within 1.75 V - 4.0 V. The ASSB full cell assembly requires extensive optimization of the fabrication process, which is out of the scope of this study focusing on developing new CCPOs as solid electrolytes (as the model system to validate new materials discovery strategies via CCCs utilizing non-equimolar compositional designs, along with microstructure and interfacial engineering via processing) and should be pursued as future work. Nonetheless, the results in this study suggest that LSTNZH is promising for real applications in ASSBs.

**Conclusions**

In summary, we have discovered a new class of CCPOs as solid electrolytes with improved properties via new compositionally complex materials discovery strategies, followed by a systematic investigation on the influence of mobile carriers (Li$^+$), vacancy sites (V$_{sr}^{''}$), and B-site aliovalent substitution (Sn$^{4+}$, Nb$^{5+}$) on phase stability, grain size, and ionic conductivity. We have further discovered the composition-phase-property relationship to suggest a suitable cation stoichiometry range for maintaining single phase upon maximizing A-site carrier concentration. Using a fitted active learning MTP, MC/MD simulations revealed increased Sr vacancies and equal or greater Li$^+$ concentration at the GB regions of LSTNZS (when compared to bulk) that explained the reduced GB resistance. Moreover, a statistical analysis of the microstructures uncovered increased grain sizes of CCPOs, which reduced the amounts of GBs to improve total ionic conductivity. To unravel the origin of enhanced grain growth in CCPOs, a series of (Li$_{0.375}$Sr$_{0.4375}$)(Ta$_{0.75-z}$Nb$_z$Zr$_{0.125}$Hf$_{0.125}$)O$_{3-\delta}$ were synthesized and characterized. The controlled experiment concluded that Nb$^{5+}$ substitution in B-site promotes grain growth, thereby reducing the total (apparent) GB resistance. An additional interesting and highly useful finding was that quenching can be adopted to enhance the GB, and therefore the total, ionic conductivity of LSTNZH, thereby providing a new knob to tailor and improve the properties. Aberration-corrected STEM-EDS attributed the 77% increase in specific (true) GB ionic conductivity to compositional and structural changes in the GBs of air-quenched LSTNZH. For A-site elements at LSTNZH GBs, DFT calculations revealed that the Sr ions have higher tendency to leave the GB regions than Li ions. Along with MTP MC/MD results of LSTNZS, the calculations suggest that GB Li depletion is less a problem for these CCPOs to reduce GB specific ionic conductivity. Finally, cyclic voltammogram of LSTNZH showed stable cycling from 1.75 V to 4.0 V, with comparable electrochemical stability as the state-of-the-art LSTZ baseline, but achieving ~2.7× increase in the total ionic conductivity.

In a broader perspective, this study has established new strategies to tailor CCCs via a seamless combination of (1) non-equimolar compositional designs and (2) microstructure and interface engineering via processing. Using solid electrolytes as an exemplar, we have validated these new strategies in this study via discovering a new class of CCPOs to show the possibility of improving ionic conductivities beyond the limit of conventional doping, where controls of the microstructures and interfaces are important beyond the complex compositional designs. This work opens a new window for discovering compositionally complex ceramics or CCCs for energy storage and many other applications.

**Experimental Procedures**

**Materials Synthesis**

All specimens were synthesized using solid-state reactions. The precursor Li$_2$CO$_3$ (Acros Organics, 99.999%), SrCO$_3$ (Alfa Aesar, 99.99%), Ta$_2$O$_5$ (Inframat Advanced Materials, 99.85%), Nb$_2$O$_5$ (Alfa

Aesar, 99.9%), $ZrO_2$ (US Research Nanomaterials, 99.9%), and $SnO_2$ (US Research Nanomaterials, 99.9%) powders were used for LSTNZS series, and $HfO_2$ (Alfa Aesar, 325 mesh) was also used for LSTNZH series instead. The 10 wt.% excess $Li_2CO_3$ precursor was added for LSTNZS to compensate $Li_2O$ loss during high-temperature sintering (Figure S27). All weighted precursors were ball-milled using SPEX 8000D high-energy ball mill for a continuous 100 min. Subsequently, the powder mixture was calcinated at 800 °C for 2h in air to remove carbonates and was pressed into green pellets with 10 mm in diameter by a hydraulic press. The specimens were sintered at 1300 °C for 12h in air. The air-quenched sample was taken out from the high-temperature furnace at 1300 °C after 12h sintering and placed in air to quench to room temperature (> 40˚C/min cooling rate). All as-sintered pellets were ground and polished on both sides before material characterizations.

**Electrical Measurements**

Ionic conductivity was evaluated at 25 °C and frequencies from 100 Hz to 40 MHz with an applied voltage amplitude of 0.1 V, using a Hewlett-Packard 4194A Impedance Analyzer. To validate the negligible electronic conductivity contribution in the impedance measurement, DC polarization was measured at constant voltage of 0.1V for 8h, and measured from 0.5V to 5.5V with a 1.0V interval for 12h at each step, using a Solartron potentiostat. In Figure S28, the ionic transference number of LSTNZH ($x = 9/16$) was calculated ($t_i = 0.998$ at 0.1V), and the resistance was fitted as the slope of potential-electronic current line ($91.81 \pm 7.27$ MΩ). Before the electrical measurements, both planar surfaces of the sintered pellet were coated with lithium-ion blocking Ag electrodes.

**Crystallography and Microstructure Characterization**

The crystal structure was characterized by X-ray diffraction (XRD) with Rigaku MiniFlex (Cu Kα radiation, λ = 1.5406 Å, scan rate = 2.3 degrees/min, step = 0.01 degree). Scanning electron microscopy (SEM, FEI Apero) with energy dispersive spectroscopy (EDS, Oxford N-MAX) was performed to probe the elemental homogeneity using applied current 3.2 nA and voltage 20 kV. The relative densities of sintered pellets were calculated by the ratio of the experimental density (measured by weight and volume) and the theoretical density from XRD refinements. Electron back-scattered diffraction (EBSD) was conducted at 26 nA and 20 kV, using FEI Apreo LoVac SEM with Oxford Instruments symmetry EBSD detector. The grain size and coincidence-site lattice grain boundary (GB) analysis were performed using the inverse pole figures along the sample normal direction. The grain size and coincidence-site lattice (CSL) GB analysis were performed using software Tango.

**Electron Microscopy**

TEM specimen preparation procedures are described in the Supplemental Information. HAADF-STEM imaging, core-loss EELS, and EDS measurements were conducted using a JEOL JEM 300CF operated at 300 kV. The microscope was equipped with double aberration correctors, Gatan Image Filter Quantum with Gatan K2 Summit, and dual 100 mm$^2$ Si drift detectors (SDDs). Z-contrast HAADF-STEM imaging was performed with a probe convergence semiangle of 25.7 mrad and a large inner collection angle of 70 mrad. For core-loss EELS measurements, a dispersion of 0.5 eV per channel was used to collect the edges in the ultra-high energy loss regime (Sr-$L_{2,3}$, Ta-$M_{2,3}$, Ta-$M_{4,5}$, Nb-$L_{2,3}$, and Zr-$L_{2,3}$ edges), while 0.1 eV per channel was used to collect the O-K edge. These EEL spectra were obtained using the Gatan K2 Summit direct detection camera. The use of a direct electron detector allows for a low electron dose and minimizes irradiation damage. Edges in the low loss regime (Sr-$N_1$ and Ta-$O_{2,3}$) were obtained using the US1000 detector, with a dispersion of 0.1 eV per channel. All core-loss EEL spectra were collected using a 2.5 mm aperture and a spectrometer collection angle of 35.89 mrad. For EDS measurements, spectra were acquired using the dual 100 mm$^2$ SDDs. 75 scans (each with a 0.15 ms dwell time and 0.4 Å pixel size) in the same area at the GBs were summed.

**Electrochemistry**

The as-sintered pellet samples were ground into powders through high-energy ball milling. The as-ground powders were mixed with carbon back and PVDF binder at a ratio of 50:30:20 by weight in NMP solvent using Thinky mixer. The slurry was cast on the current collector foil (Cu/Al) with a film thickness control of 20 μm and dried at 80 ℃ under vacuum for 12h. The electrodes were punched into 10 mm diameter discs and assembled into 2032-type coin cells in glovebox with Li metal as counter electrodes, LP40 (1M $LiPF_6$ in 1:1 EC/DEC) with 10 vol.% fluoroethylene carbonate (FEC) as liquid electrolyte and Celgard 3501 as separator. CV measurements were implemented at 0.1 mV/s step, sweeping between vortex potentials for three cycles.

The Li-Li symmetrical cell were assembled in a glovebox using Swagelok-type cell with 0.7 mm thick LSTNZS pellet as the solid electrolyte layer. For wetting purpose at the interface of Li and LSTNZS pellet, 50 µL of LP40 liquid electrolyte with 10 vol.% FEC additive and 3501 Celgard separators were applied on both sides between LSTNZS and Li metal. The cell was cycled at a constant discharge and charge current density of 300 µA $cm^{-2}$ using Landt CT2001A battery tester.

**Simulations**

For both accurate and efficient simulations, we adopted a slightly modified chemical formula to represent LSTNZS ($y = 9/16$), *i.e.*, $Li_{7/18}Sr_{17/36}Ta_{1/3}Nb_{1/3}Zr_{2/9}Sn_{1/9}O_3$ with the experimentally optimized cation ratios. We utilized the active learning workflow in our study on pristine LSTZ with slight modification to fit an active learning moment tensor potential (MTP) for LSTNZS. The workflow is shown in Supplemental Figure S30 and described in detail in the Supplemental Information. The DFT and *ab initio* molecular dynamics (AIMD) simulation procedures are also described in Supplemental Information. The convergence criteria are the same as our work in LSTZ. The active learning scheme proposed by Podryabinkin and Shapeev was used to develop an MTP that can accurately simulate both bulk and GB structures.[61,63] An extrapolation grade $\gamma$ is defined to evaluate the extent to which a given configuration is extrapolative with respect to those in the training set, thereby correlating the prediction error without *ab initio* information. All training, active learning, evaluations, and simulations with MTP were performed using MLIP,[65,67] LAMMPS,[69] and the Materials Machine Learning (maml) Python package. The visualization of bulk and GB models were conducted using OVITO. The analysis on MD trajectories to extract diffusivities were performed with the pymatgen-analysis-diffusion package.

**Acknowledgment**

This research was supported by the National Science Foundation Materials Research Science and Engineering Center program through the UC Irvine Center for Complex and Active Materials (DMR-2011967). All the TEM experiments were conducted using the facilities at the Irvine Materials Research Institute (IMRI) in University of California, Irvine. The XRD, SEM, and EBSD experiments were performed at UC San Diego Nanoengineering Materials Research Center (NE-MRC). The computing resources are provided by the National Energy Research Scientific Computing Center (NERSC), a U.S. Department of Energy Office of Science User Facility at Lawrence Berkeley National Laboratory and the Extreme Science and Engineering Discovery Environment (XSEDE), which is supported by National Science Foundation grant number ACI-1548562.

**Author Contributions**

J.L. conceived the idea and formulated the overall research plan. J.L., X.P., and S.P.O. supervised different aspects of the project. S.T.K., T.L., and J.Q. designed the specific studies. S.T.K. developed the composition optimization strategy and the systematic composition investigation with the help of D.Z.. S.T.K. synthesized CCPO materials and implemented XRD, SEM-EDS, EBSD characterization, and electrical measurements with data analysis. S.T.K. and W.C.T. conducted electrochemical measurements. T.L., S.S., and Z.W. carried out AC impedance measurements and TEM specimen preparation. T.L. and X.W. performed STEM experiments and data analysis. J.Q. implemented MTP fitting, DFT, MC/MD simulations, and grand potential phase diagram calculations. W.T.P. constructed GB models and performed DFT calculations. S.T.K., T.L., J.Q., D.Z., and W.T.P. wrote the initial manuscript draft and J.L. revised and finalized the manuscript. All authors contributed to the discussion and revision of the manuscript.

**Declaration of Interests**

The authors declare no competing interests.

**Table 1.** Chemical composition, sintering condition, and total ionic conductivity for the optimized LSTNZS, LSTNZH, LSTZ in this work, benchmarked with LSTZ-related compounds and LLTO from the literature. All specimens, except for one labeled "air-quenched", were furnace-cooled.

| Composition | Sintering condition | $\sigma_{total}$ (mS/cm) | Reference |
|---|---|---|---|
| $(Li_{0.332}Sr_{0.5})(Ta_{0.334}Nb_{0.347}Zr_{0.211}Sn_{0.108})O_{3-\delta}$ | 1300˚C, 12h | 0.218 | This work |
| $(Li_{0.375}Sr_{0.4375})(Ta_{0.375}Nb_{0.375}Zr_{0.125}Hf_{0.125})O_{3-\delta}$ | 1300˚C, 12h<br>Air-Quenched | 0.151<br>0.256 | This work |
| $(Li_{0.375}Sr_{0.4375})(Ta_{0.75}Zr_{0.25})O_{3-\delta}$ | 1300˚C, 12h | 0.094 | This work |
| $(Li_{0.375}Sr_{0.4375})(Ta_{0.75}Zr_{0.25})O_{3-\delta}$ | 1300˚C, 15h | 0.08 | Chen et al.[26] |
| $(Li_{0.375}Sr_{0.4375})(Nb_{0.75}Zr_{0.25})O_{3-\delta}$ | 1200˚C, 15h | 0.02 | Yu et al.[43] |
| $(Li_{0.375}Sr_{0.4375})(Ta_{0.75}Sn_{0.25})O_{3-\delta}$ | 1450˚C, 15h | 0.046 | Thangadurai et al.[45] |
| $(Li_{0.375}Sr_{0.4375})(Nb_{0.75}Sn_{0.25})O_{3-\delta}$ | 1450˚C, 15h | 0.00182 | Thangadurai et al.[45] |
| $(Li_{0.34}La_{0.51})TiO_{3-\delta}$ | 1350˚C, 6h | 0.02 | Inaguma et al.[46] |

**Table 2.** Local composition of the GB regions in the four GB models before and after 5 ns MC/MD simulations at 1573 K. The atomic percentage of pristine LSTNZS ($y$ = 9/16) was listed for reference.

| Elements | Sr | Li | Zr | Ta | Nb | Sn | O |
|---|---|---|---|---|---|---|---|
| pristine | 9.7% | 8.0% | 4.6% | 6.9% | 6.9% | 2.3% | 61.7% |
| simple twist $\Sigma$3[110](110) | | | | | | | |
| MC/MD 0ns | 9.8% | 9.1% | 4.2% | 7.0% | 6.6% | 3.2% | 60.1% |
| MC/MD 5ns | 8.5% | 10.1% | 4.7% | 6.3% | 6.9% | 3.0% | 60.4% |
| symmetric tilt $\Sigma$3[110](1$\bar{1}$1) | | | | | | | |
| MC/MD 0ns | 9.1% | 8.8% | 4.6% | 7.6% | 6.9% | 2.5% | 60.6% |
| MC/MD 5ns | 7.0% | 11.1% | 4.8% | 6.8% | 7.1% | 2.0% | 61.1% |
| simple twist $\Sigma$5[100](100) | | | | | | | |
| MC/MD 0ns | 9.5% | 7.3% | 4.4% | 7.8% | 6.4% | 2.7% | 61.9% |
| MC/MD 5ns | 6.8% | 11.2% | 5.6% | 5.4% | 7.4% | 2.0% | 61.5% |
| symmetric tilt $\Sigma$5[100](0$\bar{1}$2) | | | | | | | |
| MC/MD 0ns | 9.7% | 8.1% | 4.9% | 6.4% | 6.2% | 2.9% | 61.7% |
| MC/MD 5ns | 8.2% | 8.4% | 3.0% | 9.2% | 6.0% | 2.5% | 62.6% |

# Figure Captions:

**Figure 1. Overview of several generations of CCPOs discovered and refined in this work, benchmarked with the LSTZ baseline.** The first generation CCPO LSTNZS (square) shows a total ionic conductivity of 0.218 mS/cm, which is ~2.3X of the 0.094 mS/cm of the LSTZ baseline (star), but the electrochemical stability of the Sn-containing LSTNZS is compromised. The second generation Hf-containing (and Sn-free) CCPO LSTNZH (grey circle) shows comparable electrochemical stability with LSTZ and the total ionic conductivity of 0.151 mS/cm, which can be further enhanced through quenching (red circle) to 0.256 mS/cm. The improvement via quenching is primarily due to the increase of specific GB conductivity (by 77%, albeit the bulk conductivity is also improved moderately by 26%). Overall, the total conductivity of the air-quenched LSTNZH is ~2.7X of that of the LSTZ baseline.

**Figure 2. XRD patterns and crystal structure of new series of CCPO LSTNZS and LSTNZH.** (A) The crystal structure model obtained from XRD refinement of LSTNZH ($x$ = 9/16). The A sites are occupied by Li and Sr cations, while the B-sites are occupied by Ta, Nb, Zr and Sn/Hf cations. (B) The atom arrangement in {100} plane family of LSTNZH ($x$ = 9/16) bulk along <100> under STEM HADAAF imaging (scaler bar = 1 nm). The orange dots indicate A sites, and the blue dots represent B sites. XRD patterns of baseline LSTZ and CCPOs: (C) the LSTNZS $x$ series and (D) the LSTNZH $x$ series, where the stoichiometry controlled by a general formula $Li_{(2/3)x}Sr_{1-x}(5B)_{(4/3)x}(4B)_{1-(4/3)x}O_{3-\delta}$, and (E) the LSTNZS $y$ series with optimized B-site cation ratio (fixed) and changing A-site cation stoichiometry following a formula $Li_{(2/3)y}Sr_{1-y}Ta_{0.3334}Nb_{0.347}Zr_{0.211}Sn_{0.108}$.

**Figure 3. Plots of secondary phase fraction and bulk ionic conductivity for the LSTNZS $y$ series [$(Li_{2/3y}Sr_{1-y})(Ta_{0.334}Nb_{0.347}Zr_{0.211}Sn_{0.108})O_{3-\delta}$] and the LSTNZH $x$ series [$(Li_{2/3x}Sr_{1-x})(Ta_{2/3x}Nb_{2/3x}Zr_{0.5(1-4/3x)}Hf_{0.5(1-4/3x)})O_{3-\delta}$] and simulated Li-ion diffusivities in bulk and two grain boundary (GB) models for LSTNZS.** (A) and (B) indicate the areal fraction of the secondary phases, quantified from elemental EDS maps, as function of $y$ or $x$ values (bottom x axis) and the molar ratio of Li/Sr (top x axis). (C, D) Correlation plots of fitted $\sigma_{bulk}$ with $y$ or $x$ values. The $\sigma_{bulk}$ is optimized when $y$ or $x$ = 9/16 or 0.5625 for both the Sn-containing LSTNZS $y$ series and the Hf-containing LSTNZH $x$ series. The total conductivities of two series are benchmarked with the LSTZ baseline (star). Insets are Nyquist plots. (E) The Arrhenius plot of Li diffusivity calculated using the bulk model and two selected LSTNZS ($y$ = 9/16) GBs equilibrated with MC/MD simulations at 1573 K. (F) The corresponding $D_{Li, 300K}$ (right y axis) and activation energies (left y axis).

**Figure 4. The inverse pole figures of EBSD, grain size statistics of LSTZ, LSTNZS ($y$ = 9/16), and LSTNZH ($x$ = 9/16), and the correlation plots of mean grain size and total ionic conductivity with the $Nb^{5+}$ occupancy.** The inverse pole figures along the normal direction of (A) LSTZ, (B) LSTNZS ($y$ = 9/16), and (C) LSTNZH ($x$ = 9/16) pellets. The corresponding grain size distribution histograms of (D) LSTZ, (E) LSTNZS ($y$ = 9/16), and (F) LSTNZH ($x$ = 9/16), extracted from (A)-(C), respectively. (G) The correlation of mean grain size (top left y axis), relative density (top right y axis), phase stability (single phase with bule shade; not a single phase with yellow shade), and total ionic conductivity (bottom left y axis) with the $Nb^{5+}$ occupancy [$z$ in $(Li_{0.375}Sr_{0.4375})(Ta_{0.75-z}Nb_zZr_{0.125}Hf_{0.125})O_{3-\delta}$] in B site.

**Figure 5. MC/MD simulated grain boundary (GB) structures of LSTNZS ($y$ = 9/16) based on active learning MTP.** (A) Four GB structures of LSTNZS ($y$ = 9/16) before MC/MD simulations were constructed with large-scale GB models with over 10,000 atoms. Elements on A- and B-sites are randomly assigned. The exact geometric information of those GB models was provided in Supplemental Table S7. (B) The evolution of the atomic percentage of cations at GB regions of the four GB models during the MC/MD simulations at 1573 K.

**Figure 6. The effects of the cooling rate on structure-property relationship of LSTNZH ($x$ = 9/16) samples, including GB structure (segregation) and conductivity.** (A) AC impedance spectra of furnace-cooled (gray curve) and air-quenched (red curve) LSTNZH ($x$ = 9/16) samples measured at 25 °C. Equivalent circuit model used to fit the data is shown in (A). (B) Bulk, GB, and total resistivity of furnace-cooled and air-quenched LSTNZH ($x$ = 9/16) samples. (C) HAADF-STEM image and (D) elemental maps of Sr, Ta, Nb, Hf, Zr, and O collected at the furnace-cooled LSTNZH ($x$ = 9/16) general GB. (E) HAADF-STEM image and (F) elemental maps of Sr, Ta, Nb, Hf, Zr, and O collected at the air-quenched LSTNZH ($x$ = 9/16) general GB. Scale bars are 1 nm. Additional examples are given in Supplemental Information to show the statistical significance of the observations.

**Figure 7. GB models and the corresponding DFT calculations for the stabilization energy investigation.** (A) and (B) present the lowest-energy relaxed Σ5 [100](012) and Σ3 [110](111) structures of LSTNZH, respectively. The red rectangles indicate the GB regions in the model systems. In (A), the black circle and arrow indicate a Sr locating at GB to be placed at a nearby vacancy away from GB center for calculating the stabilization energy (the driving force of elemental depletion). (C) The result of stabilization energy calculations in box plots. The squares indicate the mean values, and the black diamond represents an outlier.

**Figure 8. The electrochemical stability testing by cyclic voltammogram using half cells for LSTNZS, LSTNZH, or LSTZ.** (A) The comparison of CV curves among three sample groups, as indicated: LSTNZS (green), LSTNZH (blue), and LSTZ (grey). The half cells were assembled using Li metal (anode), LP40 with 10 vol% FEC (liquid electrolyte), and testing sample coated on Cu foil (cathode). The CV measurements were scanned within the potential window of 0 V - 3.5 V, with a scan rate of 0.1 mV/s. The enlarged region shows the onset reduction potential at 1.4 V. (B) shows the Li-Li symmetrical cell testing using LSTNZS as the solid electrolyte separator with trace LP40 liquid electrolyte and 3501 Celgard separators on both sides between LSTNZS and Li metal. The cell was tested at 300 µA $cm^{-2}$ for 120 h, with an overpotential of 200 mV. (C) Cyclic voltammogram of half-cell assembled using Li metal (anode), LP40 with 10 vol% FEC (liquid electrolyte), and LSTNZH coated on Al foil (cathode). The CV curves were scanned from 0.8 V to 4.3 V at a scan rate of 0.1 mV/s.